\documentstyle[12pt,aaspp4,eqsecnum]{article}

\slugcomment{
\vbox{
{\it The Astrophysical Journal}, accepted.}}

\lefthead{Varadi et al.}
\righthead{Random Lag SCSA}

\newcommand{\be}{\begin{equation}}
\newcommand{\ee}{\end{equation}}

\newcommand{\msp}{{\ \ \ \ }}

\begin{document}
\title{RANDOM LAG SINGULAR CROSS-SPECTRUM ANALYSIS}

\author{
F. Varadi\altaffilmark{1},
R. K. Ulrich,
L. Bertello,
and
C. J. Henney}
\affil{Department of Physics and Astronomy,
        Division of Astronomy, \\
        University of California, Los Angeles, CA 90095-1562}

\altaffiltext{1}{also Institute of Geophysics and Planetary Physics,\\
        University of California, Los Angeles, CA 90095-1567}

\begin{abstract}
In a previous paper (\cite{varadi99}), Random Lag
Singular Spectrum Analysis was offered as a tool to find
oscillations in very noisy and long time series.
This work presents a generalization of the technique
to search for common oscillations in two or more
time series.

\end{abstract}


\section{INTRODUCTION}

This paper follows up on our previous one (\cite{varadi99}) 
in which Random Lag Singular Spectrum Analysis was described
at length. The technique is being used to search for
low-frequency solar acoustic and gravity oscillations in 
the Global Oscillations at Low Frequency (GOLF; \cite{gab}) and 
Michelson Doppler Imager (MDI; \cite{sch}) data.
It has became apparent, however, that identifying
common, simultaneous oscillatory components in these data
is a more promising approach.
Hence the need arose to generalize the technique for two or more
time series which is described here briefly.

Singular Spectrum Analysis (SSA; \cite{ssa89}; \cite{ssa92}; \cite{ssa95}) 
was originally developed to search for
oscillation in short and noisy time series that one typically
encounters in geophysics. 
The technique computes the eigenvectors of autocorrelation
matrices. The sizes of the latter are usually a third
of the length of the time series which
limits the feasibility of the method to time series no longer
than a few thousand. 
It has been employed in the analysis of GOLF and MDI
data by extracting the signal in narrow frequency bands which reduces
the length of the time series to which SSA is applied
(\cite{ulrich98}; \cite{varadi98}).
This, however, made it difficult to assess the importance 
of candidate modes in the signal as a whole.
Our previous paper on Random Lag Singular Spectrum Analysis 
(\cite{varadi99}) explored connections between linear dynamical 
systems and SSA. It was shown that one can work with sequences
of random lags when dealing with matrices of autocovariances,
which are the mainstay of SSA and numerous other
techniques such as autoregressive modeling (AR)
(e.g., \cite{percival93}; \cite{proak}).
This approach makes it possible to carry out SSA in wide frequency bands,
on time series having tens of thousands of points.
Here a generalization of the technique,
Random Lag Singular Cross-Spectrum Analysis (RLSCSA), 
is introduced through a few equations, 
without exploring deep mathematical issues.
Preliminary results obtained by RLSCSA are quite
encouraging and it seems appropriate to make the technique
accessible to the community expeditiously.

\section{RANDOM LAG SINGULAR CROSS-SPECTRUM ANALYSIS}

The technique searches for coincident patterns
of oscillations in two given time series, 
$x = x_1, x_2, \ldots, x_N$ 
and $y = y_1, y_2, \ldots, y_N$ with the same uniform sampling in time.
Random Lag Singular Spectrum Analysis (\cite{varadi99}) 
does so in the case of a single time series, i.e, $x=y$, 
by looking at possible linear relationships between its lagged copies.
The reasoning relies on straightforward
but lengthy linear algebra.
In short, a linear system produces time series 
in which consecutive values are linear related, at least approximately.
This is analogous to the point of view of AR
models (\cite{percival93}) but it is more general. 
The most important difference, however, 
is that SSA and its generalizations do 
not try to fit an AR model to very noisy data
but rather try to extract what appears to be 
signal from a noisy background.

Both time series are assumed to have zero mean.
First two matrices are formed, one of which is
\begin{equation}
{\bf D}_x = 
\left(
\begin{array}{cccc}
\vdots & \vdots & & \vdots \\
x_{j-1+k_1} & x_{j-1+k_2} & \ldots & x_{j-1+k_M} \\
x_{j+k_1} & x_{j+k_2} & \ldots & x_{j+k_M} \\
x_{j+1+k_1} & x_{j+1+k_2} & \ldots & x_{j+1+k_M} \\
\vdots & \vdots & & \vdots \\
\end{array}
\right) \; ,
\end{equation}
whose columns are index-shifted (lagged) copies of the time series $x$.
Here the lags $k_1, k_2, \ldots, k_M$ are all different, random positive integers
uniformly distributed between 1 and a maximum lag $K$.
With some other set of random lags, the analogous matrix
${\bf D}_y$ is formed for $y$. The two sets of lags, for $x$ and $y$,
may or may not contain common elements.
For the sake of exposition,
the data matrices ${\bf D}_x$ and ${\bf D}_y$
are padded with zeros for those indeces of $x$ and $y$ 
for which no value is available, i.e., when the index
is smaller than one or larger than $N$.
In practice, these data matrices are not used directly.
Then the $M\times M$ matrix 
\begin{equation}
{\bf C} =
{\bf D}^{T}_x {\bf D}_y 
\end{equation}
is formed which consists of covariances at various lags, except for
normalization factors ($^T$ denotes transpose).

The essence of the method is to compute the Singular Value Decomposition
(SVD; \cite{gol}) of this matrix, i.e.,
\begin{equation}
{\bf C} = {\bf D}^{T}_x {\bf D}_y = 
{\bf E}_x {\bf \Lambda} {\bf E}^{T}_y ,
\label{cceq}
\end{equation}
where 
${\bf E}_x$  and ${\bf E}_y$ 
are orthogonal matrices. The matrix ${\bf \Lambda}$ is diagonal and contains
the so-called singular values.
From (\ref{cceq}) it follows that
\begin{equation}
\left({\bf D}_x {\bf E}_x \right)^T
\left({\bf D}_y {\bf E}_y \right) = {\bf \Lambda}.
\end{equation}
The columns of the $N\times M$ matrix ${\bf D}_x {\bf E}_x$ represent 
filtered versions of the original time series $x$, while
${\bf D}_y {\bf E}_y$ does the same for $y$.
Clearly, the singular values can be interpreted as covariances
between these filtered time series.
One can also observe that the $i$th column of
${\bf D}_x {\bf E}_x$ has nonzero covariance only with
the $i$th column of ${\bf D}_y {\bf E}_y$.

Next the columns of ${\bf D}_y {\bf E}_y$ are modeled
by the columns of ${\bf D}_x {\bf E}_x$ using ordinary linear 
regression, the $i$th column in the former by the $i$th column
in the latter. For this, the variances of the columns in both matrices
have to be computed. In the case of $x$, they are the diagonal elements
of the $M\times M$ matrix
\begin{equation}
\left({\bf D}_x {\bf E}_x\right)^{T} 
\left({\bf D}_x {\bf E}_x\right)
=
{\bf E}^T_x \left({\bf D}^{T}_x {\bf D}_x \right) {\bf E}_x 
\label{autocovs}
\end{equation}
and an analogous formula applies in the case of $y$.
Since the original $x$ and $y$ have zero means, the same is true
for the columns of ${\bf D}_x {\bf E}_x$ and ${\bf D}_y {\bf E}_y$,
at least approximately in the case of large $N$ and no trends in the data.
Hence one needs to compute only the scaling coefficients
in the linear regression models
between these columns. These can be collected to form
a diagonal matrix ${\bf B}_x$ to obtain
\begin{equation}
\widehat{\left({\bf D}_y {\bf E}_y\right)}  = {\bf D}_x {\bf E}_x {\bf B}_x,
\end{equation}
where $\widehat{\;\;\;}$ signifies that this is a statistical model.
Next one defines the model for the columns of ${\bf D}_y$ as
\begin{equation}
\widehat{{\bf D}_y} = {\bf D}_x {\bf E}_x {\bf B}_x {\bf E}^T_y.
\end{equation}
Finally, we have to create a new time series $\widehat{y}$ from $\widehat{{\bf D}_y}$.
Each column of the latter is index-shifted 
relative to the original indexing of $y$
and thus one could use any column as $\widehat{y}$.
Alternatively, one can simply unshift each column and average them 
to obtain
\begin{equation}
\widehat{y}_n = \frac{1}{M} 
\sum_{s=1}^M
\sum_{i=1}^M 
\sum_{j=1}^M 
x_{n+(k^x_i-k^y_j)}
\left(B^{s}_{x}\right)_s
\left(E^{s}_{x}\right)_i
\left(E^{s}_{y}\right)_j ,
\label{filtereq}
\end{equation}
where $k^x_i$ is the $i$th lag for $x$,
$k^y_j$ is the $j$th lag for $y$,
$s$ designates the $s$th column of a matrix and the subscripts $i$ and $j$
are row indices. 

The last equation describes how to model oscillations
in $y$ with those in $x$ by moving-average or finite 
impulse response filters.
Very roughly, the oscillations in $x$ are represented by
${\bf E}_x$ and those in $y$ by ${\bf E}_y$, while
${\bf B}_x$ contains scaling factors.
At this point one can separate signal and noise
by not taking into account all the linear regression models
between $x$ and $y$. When the
cross-correlation is large for some component $s$,
it is more likely that the same oscillation is present
in both time series. Hence one would not sum for all 
$s = 1, \ldots M$ but for only some of them. 
Once the filter has been determined, perhaps the best
way to proceed is by computing its spectral response
i.e., its $z$ transform
which boils down to computing its Fourier transform (\cite{proak}).
This is analogous to the Maximum Entropy Method (\cite{percival93}).

When $x$ and $y$ are the same and the same lags are used,
the filtering formula above is the same as in Random Lag Singular Spectrum
Analysis with ${\bf B}_x$ being identity matrix
(\cite{varadi99}). Therefore, RLSCSA is a direct generalization
of Random Lag Singular Spectrum Analysis. 
Furthermore, as in the case of the latter, in RLSCSA 
the analysis can be carried out for several different lag sequences 
and the filters can be averaged for these cases. 
Also, a larger number of lags, $M$, always provides better results.

There can be a number of variations on the basic construction
above. For instance, in the case of $x=y$ one can use
lag sequences $k^x$ and $k^y$ which do not have common
elements. This doubles the number of autocovariances one would include
in the matrix ${\bf C}$ as compared to Random Lag Singular Spectrum Analysis
for the same number of lags $M$.
In the case of $x\neq y$, one could use the same lag sequence
($k^x = k^y$) or different ones.
When several time series are given, divided into two groups,
$x^{(1)}, x^{(2)}, \ldots, x^{(d_x)}$
and
$y^{(1)}, y^{(2)}, \ldots, y^{(d_y)}$,
one can form the data matrix
\begin{equation}
\left[
{\bf D}_{x^{(1)}},
{\bf D}_{x^{(2)}},
\ldots
{\bf D}_{x^{(d_x)}}
\right]
\end{equation}
and an analogous one for the other set of time series.
In these matrices the columns come from possibly different time series.
The formulae above needs only straightforward modifications for this case
although indexing becomes somewhat complicated.
In (\ref{filtereq}), one would have filters
operating on the $x$ signals whose sum
would model oscillations in $y^(i)$ for $i=1,2,\ldots, d_y$.

There can be phase lags between $x$ and $y$.
There should be none, however, when they are
simultaneous and independent measurements of the same 
physical quantity.
In such cases, one can use the phase information
in the filter (\ref{filtereq}) to further separate signal and noise.
If ${\bf E}_x$ and ${\bf E}_y$
operate on the same signal, then the phases
obtained should be close to each other.
Hence one would consider as signal
only those peaks in the spectral
response of the filter which have nearly zero phase.
Such a ``phase weeding" seems to provide better results
but further work is needed to fully develop this idea.

\section{IMPLEMENTATION}

For clarity, we provide a description of how the
computations are done in practice.
Instead of products of lagged time series,
estimates of covariances are used to create the matrix ${\bf C}$.
First of all, we subtract from $x$ its mean to make sure that it is zero and
also devide the resulting $x$ by its standard deviation
to ensure that the numerical values are in a reasonable range.
The same is done for $y$.
Next one has to determine cross- and autocovariances
which, in turn,  are computed
by fast convolution algorithms (e.g, \cite{percival93}; 
\cite{proak}; \cite{varadi99}).
For that, the quantities
\be
\widehat{q}_j = \sum_k x_k y_{k+j},  \msp j = -(K-1), \ldots, K-1 \, 
\ee
are computed the following way.
If $\tilde{y}$ denotes the reverse of $y$, i.e.,
\be
\tilde{y}_i = y_{N-i+1},
\ee
and the number $N_2$ is at least $2N-1$,
then
\be
\widehat{q} = \frac{1}{N_2} {\rm DFT}_{N_2}^{-1}
\left( {\rm DFT}_{N_2}(x)
\,
{\rm DFT}_{N_2}(\tilde{y})\right).
\label{conform}
\ee
Here ${\rm DFT}_{N_2}(x)$ denotes the complex Fourier transform on
$N_2$ points without normalization, i.e., without dividing
the transform with $N_2$, in either the direct or inverse transform.
In formula,
\be
 {\rm DFT}_{N_2}(z)_k = \sum_{j=0}^{N_2-1} z_j
e^{-2\pi jk/N_2},  \msp
 {\rm DFT}^{-1}_{N_2}(z)_j = \sum_{k=0}^{N_2-1} z_k
e^{2\pi jk/N_2} .
\ee
Both $x$ and $\tilde{y}$ are first index-shifted to start with
zero index and then are padded by zeros to length $N_2$. The latter
ensures that the otherwise cyclic (or circular) convolution obtained by
(\ref{conform}) is in fact acyclic (e.g., \cite{proak}).
After computing the right hand side of (\ref{conform}),
the result has to be index-shifted by $N$ in the negative
direction to obtain the values of $\widehat{q}$ with the correct indexing
since the $N$th element on the right hand side is actually $\widehat{q}_0$.
Next we compute
\be
q_j = \frac{1}{N-j} \widehat{q}_j, \msp j = -(K-1), \ldots, K-1,
\ee
which are unbiased estimates of the covariances --- which can proved
along the same lines as for autocorrelations (\cite{proak}).

Next the $M$-long sequences of random lags are selected, $k^x$ and $k^y$,
using a random number generator.
Having the latter, one can form the matrix
${\bf C}$, more exactly its multiple by some factor.
One has
\be
C_{ij} = q_{k^y_j - k^x_i}.
\ee
The SVD of ${\bf C}$ is computed the following way but there are 
other methods (\cite{gol}).
First one computes the symmetric matrix  
\begin{equation}
{\bf C}^T
{\bf C} = 
{\bf E}_y
{\bf \Lambda}^2
{\bf E}^T_y
\end{equation}
whose eigenvectors are the columns of ${\bf E}_y$.
Any standard high-performance algorithm for the symmetric
eigenvalue problem (\cite{gol}) can be used to compute ${\bf E}_y$.
For the matrix ${\bf E}_x$ one has
\begin{equation}
{\bf C} {\bf E}_y = {\bf E}_x {\bf \Lambda},
\end{equation}
which means that ${\bf E}_x$ can be computed
by normalizing the columns of the matrix on the left hand side.
The normalizing factors are the singular values which
are all nonnegative if these formulae are used.
One has to be careful when very small singular values
are present but this did not occur in the cases we dealt
with so far. 

The autocovariances in (\ref{autocovs}) are computed
as above with $x$ replacing $y$. 
The diagonal matrix of scaling coefficients, ${\bf B}_x$, 
is determined by the standard formula
\be
B_k = \frac{\Lambda_k}{\sigma^2_k}
\ee
where the index $k$ refers to the $k$ element in the
diagonals and $\sigma^2_k$ is the $k$th element
in the diagonal of (\ref{autocovs}).
Finally, the filters are computed using (\ref{filtereq}).
In this, one has to make choices which components should
be included, i.e., for which values of $s$ should the
sum be restricted. Obviously, one should compute
the correlation coefficients
\be
r_k = \frac{\Lambda_k}{\sigma^{x}_k \sigma^{y}_k}
\ee
where the superscript of $\sigma$ is used
to distinguish between the standard deviations 
of the $k$th columns of
${\bf D}_x {\bf E}_x$ and ${\bf D}_y {\bf E}_y$.
Then the filter coefficients in (\ref{filtereq}) are
collected for those $s$ for which $r_s$ are the largest.
The computational steps starting with the selection
of random lags are repeated for a number of
different selections of them and the resulting
filter coefficients are averaged, as in the case
of Random Lag Singular Spectrum Analysis (\cite{varadi99}).

\section{AN EXAMPLE}

It is worthwhile to present a ``proof of concept" example.
A second-order autoregressive model, i.e.,
a discrete, damped oscillator,
was used to generate a signal by forcing the model with
white noise. The signal has 50000 points from which 
two subsequences were selected, each 5000 long, separated
from each other by 40000. 
In the top panel of Fig. \ref{fig1}, the Fourier spectrum of one 
of the subsequences is shown. 
Then white noise was added to both subsequences.
As can be seen in the middle panel of Fig. \ref{fig1}, the 
cross-spectrum amplitudes of the two noisy subsequences reveal nothing
about original signal. The bottom panel shows the 
spectral response of the RLSCSA filters. 
Here 500 lags were used, the maximum lag being 2500, and 
the filters in (\ref{filtereq}) were averaged for 100 different 
sets of random lags. The components with largest two correlations
were included in the filter for each selection of lags.
The signal is recovered as the third largest peak
in the spectral response which we find
quite encouraging. This example also illustrates that 
one should expect to see a number of noise peaks beside signal
peaks. Still, one has to deal with a few peaks in the RLSCSA results,
while the cross-spectrum exhibits hundreds of them.

\acknowledgments
We thank M. Dettinger, I. Fodor, M. Ghil, J. Leibacher, J. Pap, 
C. Tatsuoka and P. Yiou for enlightening discussions on time series 
analysis and statistics. 
SOHO is a project of international cooperation
between ESA and NASA.
This research is supported by a NASA subcontract
to UCLA through Stanford University.
The partial financial support of NSF Grant AST96-19574 and
NASA Grant NAG5-6680 (F.V.) is also acknowledged.

\pagebreak

\count255=0
\def\numit{\advance\count255 by 1 \vfill\hfill Fig. \number\count255}

\def\getplot#1{ \vbox to 7in { \vskip 0in \hskip -1.0in
\epsfverbosetrue \epsfxsize=7in \epsfbox{Fig#1.eps} }
}

\getplot{1}

\figcaption[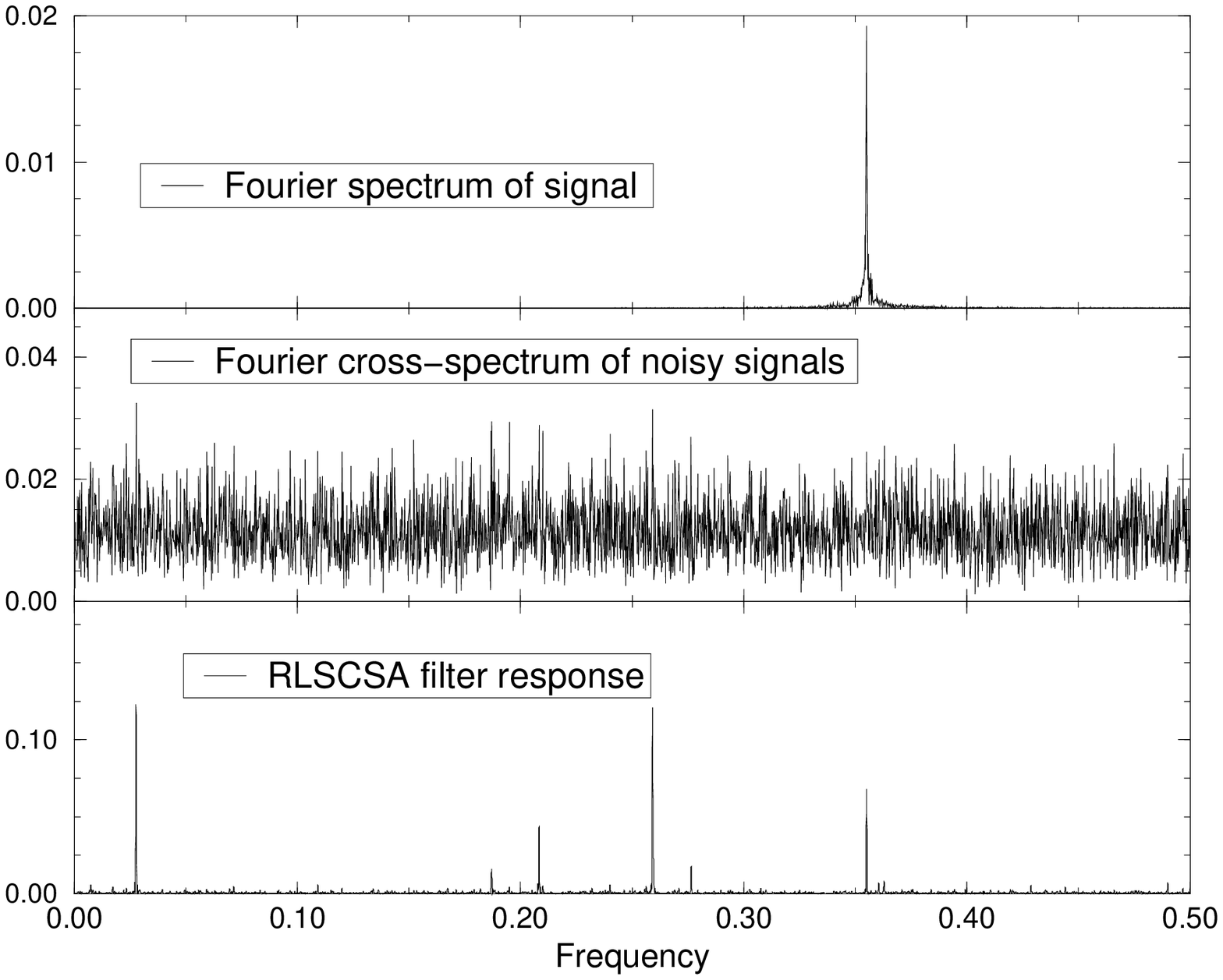]{
Top panel: Fourier spectrum of one subsequence of the signal.
Middle panel: Amplitude of the cross-spectrum of two subsequences
of the noisy signal. The original signal is completely
hidden in noise.
Bottom panel: RLSCSA filter response. The third largest peak is the
original signal.
\label{fig1}}

\end{document}